\documentclass[prb,twocolumn,preprintnumbers,amsmath,amssymb,floats,citeautoscript,showpacs]{revtex4-1}

\usepackage{graphicx}
\usepackage{color}

\begin{document}


\title{Novel Charge Ordering in the Trimer Iridium Oxide BaIrO$_3$}

\author{Ichiro Terasaki}
\email[Email me at:]{
terra@cc.nagoya-u.ac.jp
}
\author{Shun Ito}
\author{Taichi Igarashi}
\author{Shinichiro Asai}
\author{Hiroki Taniguchi}
\affiliation{Department of Physics, Nagoya University, Nagoya 464-8602,
Japan}

\author{Ryuji Okazaki}
\affiliation{Department of Physics, Faculty of Science and Technology, 
Tokyo University of Science, Noda 278-8510, Japan}

\author{Yukio Yasui}
\affiliation{Department of Physics, Meiji University, 
Kawasaki 214-8571, Japan}

\author{Kensuke Kobayashi}
\author{Reiji Kumai}
\author{Hironori Nakao}
\author{Youichi Murakami}

\affiliation{Condensed Matter Research Center and Photon Factory, 
High Energy Accelerator Research Organization, Tsukuba 305-0801, Japan}

\begin{abstract}
We have prepared polycrystalline samples of the trimer Ir oxide
BaIrO$_3$ with face-shared Ir$_3$O$_{12}$ trimers,
and have investigated the origin of the phase transition at 182 K
by measuring resistivity, thermopower, magnetization and
synchrotron x-ray diffraction. 
We propose a possible electronic model 
and transition mechanism,
starting from a localized electron picture
on the basis of the Rietveld refinement.
Within this model, BaIrO$_3$ can be basically regarded as a Mott insulator,
when the Ir$_3$O$_{12}$ trimer is identified to one pseudo-atom
or one lattice site. 
The transition can be viewed as a transition from the Mott insulator
phase to a kind of charge ordered insulator phase.
\end{abstract}


\maketitle

\section{Introduction}

Strongly correlated electrons are seen in a certain class of solids
in which the Coulomb repulsion is too strong to hold
a simple one-electron picture based on the band theory.
They have occupied a central position in condensed matter sciences 
for many decades, and 
have attracted keen interests from vast numbers of
researchers since the discovery of high-temperature
superconducting copper oxides in 1986.
They store a macroscopic number of
degeneracy and entropy based on spin and charge degrees of freedom
on each lattice site at high temperature 
\cite{furukawa1993,terasaki2002}, 
and release them through various phase transitions towards 0 K. 
Various forms of phase transitions and
ordered states  have been discovered one after another
with the progress in the studies of strongly correlated electrons.
In this context, understanding of new phase transitions and
ordered states is a major purpose for strong-correlation physics.

\begin{figure}
 \centering
 \includegraphics[scale=.35]{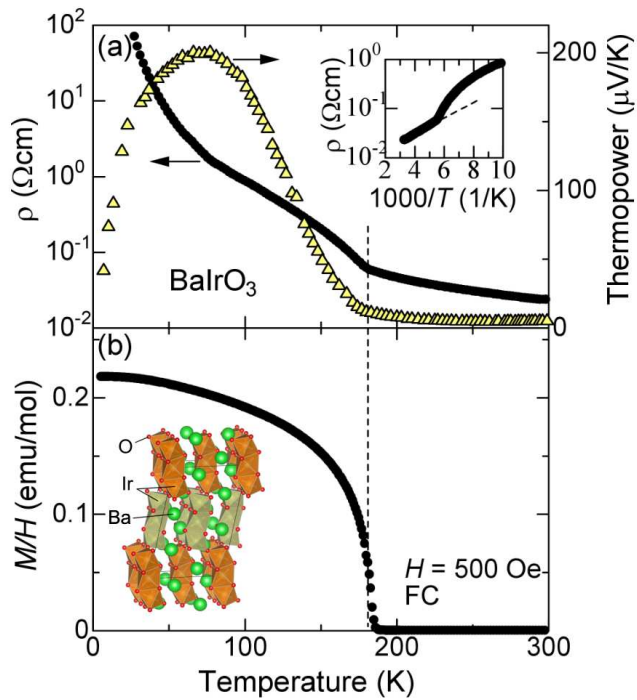}
 \caption{(a) Resistivity and thermopower
of a polycrystalline sample of BaIrO$_3$.
The inset shows the Arrhenius plot of the resistivity,
in which the dotted line indicates the activation transport
above the transition temperature $T_c$.
(b) Magnetization $M$ divided by an external field $H$
of 500 Oe of the same sample.
The dotted line represent $T_c=$182 K.
The inset shows the schematic picture of the
crystal structure of BaIrO$_3$.
}
 \label{fig1}
\end{figure}

BaIrO$_3$ ($18R$ phase)
is an interesting oxide as a playground for a new phase transition. 
The crystal structure is schematically shown in the inset of
Fig. \ref{fig1}(b),
in which the three IrO$_6$ octahedra are connected with one another
in a face-sharing network, and form an Ir$_3$O$_{12}$ trimer structure,
as indicated in light brown and dark yellow.
The trimers are connected with each other in a corner-sharing network
and construct zig-zag chains along the $c$ axis and corrugated honeycomb
lattices in the $ab$ plane.
Owing to the low crystal symmetry of $C2/m$, two trimers  are inequivalent
in the unit cell, as painted in different colors in the inset of Fig. \ref{fig1}(b).

The study of this oxide has a rather long history;
Donohue \textit{et al}. first reported 
the synthesis and structure of this material in the 60's, \cite{donohue1965},
and  later Gai \textit{et al}. \cite{gai1976} 
studied the poly-types of 5$H$ and 6$H$.
Chamberland \cite{chamberland1991}
extensively studied the relationship of the poly-types 
to the synthetic routes, 
and further found a magnetic phase transition near 200 K.
The physical and chemical properties of the poly-types have been measured
rather recently because of difficulty in
the sample synthesis \cite{cheng2009,zhao2010}.
Siegrist and Chamberland \cite{siegrist1991}
prepared single crystals and determined the crystal
symmetry of $C2/m$, which can be viewed as
pseudo-rhombohedral $R$-centered.  
Later Powell and Battle \cite{powell1993}
identified that this magnetic order is ferromagnetic,
by measuring the temperature hysteresis and the magnetization-field curve.
Linsay \textit{et al}. \cite{linsay1993} 
measured the remnant magnetization to demonstrate 
the ferromagnetic ground state.
Cao \textit{et al}. \cite{cao2000}
measured the transport and optical properties
of a single-crystal sample of BaIrO$_3$,
and found that the ferromagnetic order accompanies
rapid increase in the resistivity below $T_c\sim$180 K.
In addition, they  observed a gap-like structure in the optical conductivity
below around 1000 cm$^{-1}$,
and  attributed this transition to charge-density-wave 
(CDW) formation with ferromagnetic order.
They further found the additional two temperature anomalies
near 80 and 30 K.
Although the 80-K anomaly was ill-defined, the 30-K anomaly was
detected by other groups 
in a muon-spin-relaxation experiment \cite{brooks2005} and
in a nonlinear conduction measurement \cite{nakano2006}.
Kini \textit{et al}. \cite{kini2005} measured the specific heat of BaIrO$_3$ and
observed a small jump of 2 J/mol K around 180 K, not around 30 and 80 K.
They also found that the thermopower rapidly increases below $T_c$,
and pointed out the existence of a charge gap below $T_c$.
Maiti \textit{et al}. \cite{maiti2005}  discussed the CDW state
through photoemission spectroscopy.

In this paper, we try to address the nature and mechanism of 
the 180-K phase transition in BaIrO$_3$ on the basis of
the electronic states constructed from a localized electron picture.
A concern in the CDW scenario is that there is
no evidence for the lattice distortion at $T_c$.
In conventional CDW materials, a lattice modulation of $2k_F$
emerges below $T_c$ through the electron-phonon interaction,
where $k_F$ is the Fermi wavenumber \cite{gruner1988}.
To the best of our knowledge, no superlattice reflections
below $T_c$ have been reported in BaIrO$_3$, and this implies
that the charge gap opens without translational-symmetry breaking.
We have investigated the structure--property relationship
through synchrotron x-ray diffraction, and 
propose a possible mechanism of the transition.

\section{Experimental Section}

Polycrystalline samples of 
BaIr$_{1-x}$Ru$_x$O$_3$ ($x\le$ 0.1)
were prepared using a conventional solid-state reaction.
A stoichiometric mixture of powder sources 
of BaCO$_3$ (99.9\%), 
RuO$_2$ (99.9\%)
and Ir (99\%) was thoroughly ground in an agate mortar.
The mixture was then pre-sintered in an alumina crucible at
900$^{\circ}$C for 12 h in air, and was pressed into pellets after
re-grinding.
The pellets were sintered at 1000$^{\circ}$C for 36 h in air.

The synchrotron x-ray diffraction was taken at BL8A\&8B,
Photon-Factory, KEK, Japan.
The x-ray energy  was adjusted to be 18 keV, which was carefully
calibrated using a standard powder sample of CeO$_2$.
Powder samples were sealed in a silica-glass capillary of
0.1-mm diameter, and the capillary was rotated by an angle of
30 deg from the sample-stage axis during measurement.
The sample temperature was controlled using a cool 
helium gas or nitrogen gas. 
The diffraction patterns were analyzed using the Rietveld refinement
with Rietan-FP code \cite{FP}.

The resistivity was measured with a four-probe technique using
a home-made measurement system in a liquid-helium cryostat.
The thermopower was measured with a two-probe method 
using a home-made measurement system in a liquid-helium cryostat,
and the thermopower of the measurement leads was carefully 
subtracted. The temperature difference was typically 0.5 K
which was monitored with a copper-constantan differential thermocouple.
The magnetization was measured with a commercial product of 
Quantum Design Magnetic-Property-Measurement-System.

\begin{figure}
 \centering
 \includegraphics[scale=.20]{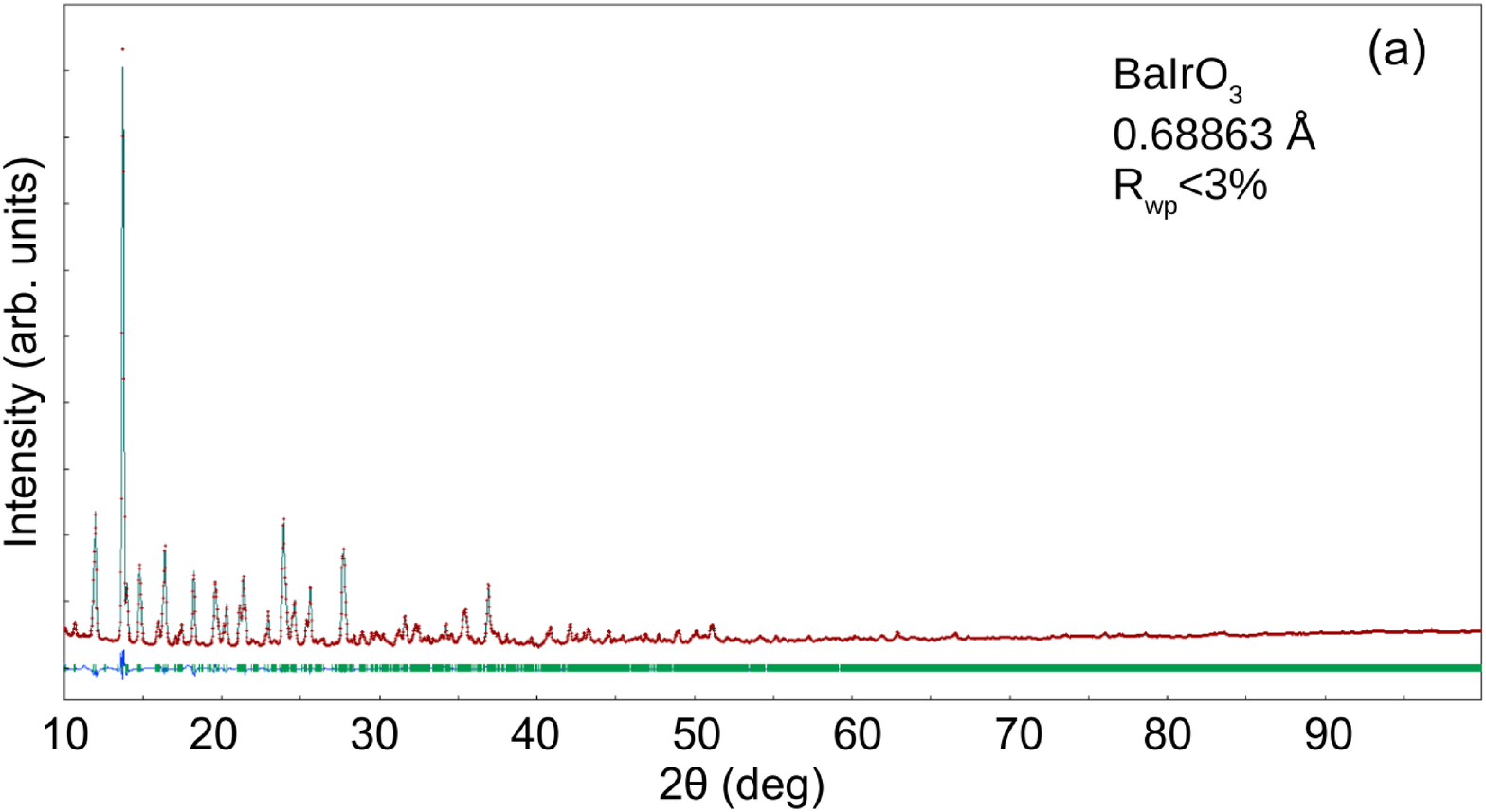}

\vspace*{6mm}
\centering
 \includegraphics[scale=.35]{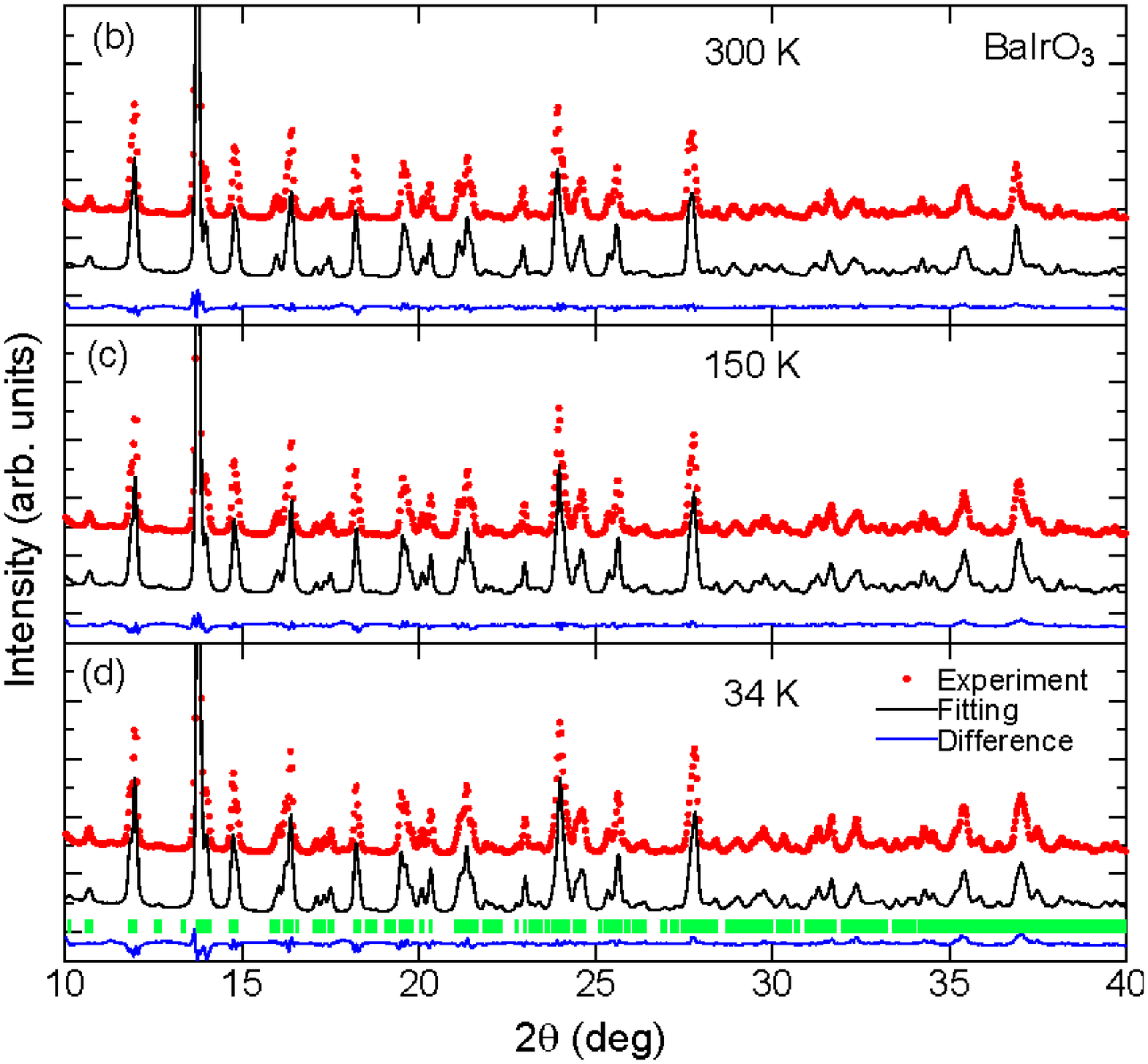}
 \caption{(a)Synchrotron x-ray diffraction pattern
of BaIrO$_3$ at room temperature. 
For the details of the refinement, see text.
Comparison between the observed data and the refinement
at (b) 300  K, (c) 150 K and (d) 34 K.} 
 \label{fig2}
\end{figure}

\section{Results}
Figure \ref{fig2} shows a synchrotron x-ray diffraction pattern
of BaIrO$_3$ at room temperature. 
Since the scattering intensity of oxygen is 100 times weaker
than that of Ba and Ir, we gave up determining the atomic position
of oxygen atoms, but instead employed the values reported
in the literature \cite{siegrist1991}.
Then we optimized the atomic positions of Ba and Ir atoms 
above 100 K by iterating the calculation until
the $R_{\rm wp}$ and $S_R$ values reached less than 3 \% and 4, 
respectively.
We accepted worse values of $R_{\rm wp}$=3 -- 4.5\% for
the data below 100 K.
The experimental data are compared with the refinement 
at 300, 150 and 34 K in Figs. \ref{fig2}(b),  (c) and (d), respectively.
As shown in Fig. \ref{fig2}, the refinement is satisfactory; 
there are neither  detectable impurity phases nor poly-type phases,
and the deviation from the refinement seems 
small enough to discuss the atomic positions
of the Ir atoms.
We measured the diffraction patterns around $T_c$ carefully,
and found no trace of superlattice reflections.
This consolidates that the phase transition does not
accompany the violation of translational symmetry,
also indicating that this is unlikely to be a CDW transition.

Figure \ref{fig1}(a) shows the transport properties of the
polycrystalline sample of BaIrO$_3$ obtained from the same batch
as used in the x-ray measurement.
The resistivity is non-metallic at all temperatures with an anomalous
increase near $T_c=$182 K. 
We should note that the resistivity above $T_c$ is of activation type
as indicated by the dotted line shown in the inset of 
Fig. \ref{fig1}(a).
The slope of the dotted line gives a rough estimate of the activation
gap $E_g^{\rho}/k_B$ around 240 K, which
is small in comparison with a band gap
of conventional semiconductors. 
This is  comparable to or even smaller than $k_BT$, 
and it is nontrivial whether or not we may regard this 
non-metallic behavior as activation-type transport.
According to the band calculation of BaIrO$_3$
\cite{whangbo2001,maiti2006,ju2013}, the electronic states
above $T_c$ are \textit{metallic} with no gap in the density of states.

The thermopower also shows an anomaly near $T_c$, below which
it rapidly enhances with decreasing temperatures.
As Kini \textit{et al.} pointed out \cite{kini2005}, 
this is a clear indication of the charge gap 
in the density of  states \cite{FeSi,klein2011},
strongly indicating an insulating nature below $T_c$.
The thermopower takes a broad maximum near 100 K, 
and goes towards zero with decreasing temperature.
We think this behavior is a crossover between 
the activation-type transport from 100 to 180 K and
the hopping conduction in the impurity states below 100 K \cite{takahashi2011}.
The thermopower above $T_c$ is  difficult to interpret;
the magnitude is as small as that of metals, but the temperature
dependence seems of activation type just above $T_c$.
We evaluate an activation energy $E_g^S/k_B$ to be 90 K
by assuming $S=E_g^S/eT$.
Again this activation energy is comparable or smaller than $k_BT$.
We will come back to this point later.

Figure \ref{fig1}(b) shows the magnetization divided by an external
field of 500 Oe measured in the field-cooling process. 
Around the same temperature of $T_c$, the magnetization 
suddenly shows up with decreasing temperature.
This ferromagnetic behavior is consistent with the previous 
reports \cite{linsay1993,cao2000,kida2008,kida2010}.
From this figure, the magnetization at 4 K is 
evaluated to be around 100  emu/mol, 
which corresponds to a saturation magnetization of 0.016$\mu_B$/Ir.
This unusually small moment is consistent with the previous 
works \cite{linsay1993,cao2000,kida2008,kida2010},
and is associated with weak ferromagnetism of itinerant magnets.
However, the electronic ground state of this oxide is
an insulator with a finite energy gap in the density of states,
and the origin of the ferromagnetism is yet to be explored.

\begin{figure}
 \centering
 \includegraphics[scale=.22]{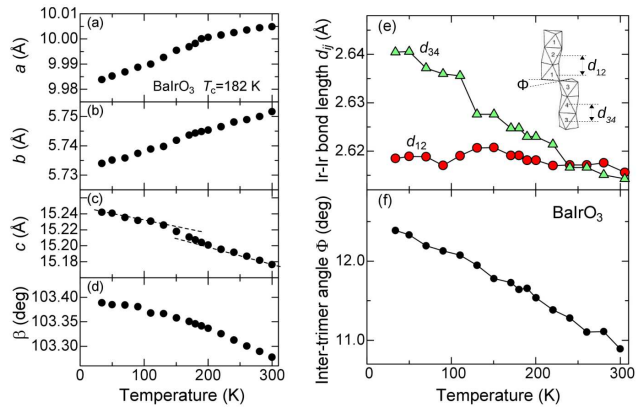}
 \caption{(a)-(d) Lattice parameters
of BaIrO$_3$. (a) $a$-, (b) $b$-, and (c) $c$-axis length.
(d) monoclinic angle $\beta$.
(e)The intra-timer Ir--Ir bond length $d_{12}$ and $d_{34}$
and (f) the inter-timer angle $\Phi$.
The inset schematically designates $d_{12}$, $d_{34}$ and $\Phi$.
}
 \label{fig3}
\end{figure}

Figures \ref{fig3} (a)-(d) show the lattice parameters plotted as a function 
of temperature obtained from the Rietveld refinement of
the x-ray diffraction patterns.
The relative uncertainty is less than 0.01\%, and the size of the 
error bars are smaller than the size of the closed circles.
All the parameters smoothly change with temperature across $T_c$,
indicating a nature of second-order transition.
By having a closer look at the data in Fig. \ref{fig3}(a),
one finds a slight cusp in the $a$-axis length near $T_c$
as a piece of evidence for finite electron-lattice coupling.

The $c$-axis length parallel to the zig-zag chain of the trimers
exhibits remarkable and nontrivial temperature dependence;
it increases with decreasing temperature as shown in Fig. \ref{fig3}(c).
Such negative temperature dependence is difficult to 
understand from solely lattice effects \cite{kittel}
except for ZrW$_2$O$_8$ \cite{mary1996}.
The negative expansion is not observed in  CDW materials
\cite{tian2000,gruner1988}, and can be another piece of evidence
that this transition is not CDW-like. 
But rather, the negative lattice expansion  
comes from various electronic origins \cite{takenaka2012}.
Prime examples are Invar alloy \cite{invar}, 
BaNiO$_3$ \cite{ishiwata2005} and
Mn$_3A$N ($A$ = Cu, Zn, Ga) \cite{takenaka2005},
in which electronic or magnetic instability tightly couples with the lattice.
The slopes of the negative expansion 
in the $c$-axis length significantly change across 
$T_c$ as shown by the dotted lines, 
clearly indicating that the negative expansion
characterizes the phase transition of the title compound. 
One may notice that the angle $\beta$ also increases with
decreasing temperature.
Since the increase in $\beta$ reduces the lattice volume, 
the $c$ axis tends to expand
in spite of the volume shrinkage  at low temperature.

In order to clarify the structural origin of the negative expansion
in the $c$-axis length, we examined what kind of atomic displacement
is responsible, and have finally arrived at the intra-trimer
Ir-Ir bond length. 
As schematically shown in the inset of Fig. \ref{fig3}(e),
there are four crystallographically inequivalent Ir sites
labeled 1, 2, 3 and 4.
We define the bond length between Ir1 and Ir2 as $d_{12}$
and that between Ir3 and Ir4 as $d_{34}$.
As shown in Fig. \ref{fig3}(e), \textit{ $d_{34}$ increases with
decreasing temperature, whereas $d_{12}$ remains constant.
This clearly indicates that the two timers are more inequivalent 
at lower temperatures.}
The data are somewhat scattered in comparison with the lattice
parameters, possibly because the latter is determined only by the
$2\theta$ values of the diffraction peaks whereas the former
needs accuracy in the relative intensity.
Nevertheless we emphasize that $d_{12}$ and $d_{34}$
were independently determined at each temperature 
within an uncertainty of 10$^{-4}$, 
and that increasing inequivalence 
at low temperatures is inherent in this oxide.
In addition, we have ascribed the structural origin for the 
temperature dependence of $\beta$ to the 
inter-timer angle $\Phi$ as indicated at the inset of
Fig. \ref{fig3}(e).
Figure \ref{fig3}(f) shows $\Phi$ plotted as a function of temperature,
where $\Phi$ increases with decreasing temperature.
We can summarize the above results as follows:
One trimer of Ir3-Ir4-Ir3 elongates with decreasing temperature,
and concomitantly the lattice distorts
to accept the elongation.

As mentioned above, 
it is difficult to determine the atomic position of the oxygen atoms
from the synchrotron x-ray diffraction used in the
present experiment, and this  constraint
prevents us from evaluating the formal valence of Ir ions
using bond-valence-sum calculation.
In the next section, we try to estimate the formal Ir valence by using
the atomic position data for oxygen from \cite{siegrist1991}.
Nevertheless we may infer from $d_{12}$ and $d_{34}$
that the average number of $d$ electrons in the
Ir3--Ir4--Ir3 timer is \textit{larger} than that in the Ir1--Ir2--Ir1 trimer,
for $d$ electrons feel the static Coulomb repulsion with one another
to stay as far as possible.
We further speculate that a finite fraction of charge rapidly transfers
at $T_c$ between the two trimers, 
and also expect that this charge transfer is a driving force 
for the phase transition.
In this respect, this charge transfer should be associated with
the CDW scenario by Cao \textit{et al.} \cite{cao2000}.
One essential difference from the conventional CDW is that
the two trimers are already inequivalent above $T_c$,
and this charge transfer causes no superlattice reflections
with holding the same unit cell  at all temperature.

\begin{figure}[t]
 \centering
 \includegraphics[scale=.25]{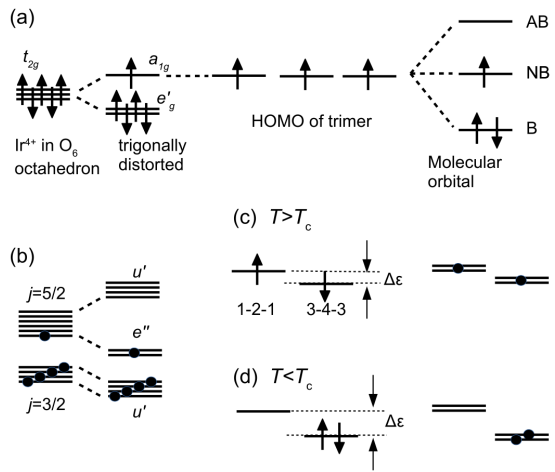}
 \caption{
A model for the electronic states of BaIrO$_3$ 
starting from a localized electron picture.
(a) The electronic configuration of Ir$^{4+}$ in a regular
and  trigonally distorted IrO$_6$ (left).
The highest occupied orbitals of Ir$_3$O$_{12}$ trimer (center)
and the molecular orbital formation (right).
B, NB and AB denote bonding, non-bonding and anti-bonding orbitals,
respectively.
(b) The electronic configuration of Ir$^{4+}$ 
in the strong spin-orbit coupling limit (taken from \cite{marco2010}).
(c) The electronic states of BaIrO$_3$ above $T_c$.
1--2--1 and 3--4--3 correspond to the inequivalent Ir$_3$O$_{12}$ trimers.
The case for strong spin-orbit coupling is drawn in the right side. 
(d)The electronic states of BaIrO$_3$ below $T_c$.
}
 \label{fig4}
\end{figure}

\section{Discussion}

Having the above results in mind, let us propose here a possible
electronic model for BaIrO$_3$ starting from the localized picture.
As schematically drawn in Fig. \ref{fig4}(a),
one Ir ion has the electronic configuration of $(5d)^5$
in the triply degenerate $t_{2g}$ level, for the formal valence of Ir is $4+$.
Since the oxygen octahedron is trigonally distorted,
the $t_{2g}$ level lifts the degeneracy to have the $a_{1g}$ 
and doubly-degenerate $e'_g$ levels.
In the present case, the octahedron is compressed 
along the (111) direction of the local coordinate, and thus 
the $a_{1g}$ level is the higher  level with one electron occupied.

There is a strong hybridization between the $a_{1g}$
levels within a trimer owing to the short intra-trimer Ir--Ir distance,
and a kind of molecular orbital is expected to be 
formed within the trimer as shown in 
the right side of Fig. \ref{fig4}(a),
where we denote
the bonding, non-bonding and anti-bonding orbitals 
made of the three $a_{1g}$ orbitals 
as B, NB and  AB, respectively. 
By putting the three electrons from the bottom,
we find one electron in the non-bonding orbital.
Thus we may regard this state as half-filling, when we
identify one trimer to one pseudo-atom.
In this sense, we propose that BaIrO$_3$ is a half-filled trimer solid.

Very recently, Ir compounds have attracted extensive interest as a
strong spin-orbit interaction system,
in which the total angular momentum of $j$ dominates the 
band structure \cite{kim2008}.
A recent x-ray absorption study has revealed
the strong spin-orbit nature in BaIrO$_3$,
and the highest occupied level should be regarded 
as the $e''$ level rather than the $a_{1g}$ level from the $t_{2g}$
levels as shown in Fig. \ref{fig4}(b) \cite{marco2010}.
Even in such a case, we can make similar molecular orbitals.
The highest occupied state then corresponds to  the NB $e''$ 
instead of the NB $a_{1g}$, and is unlikely to alter
the picture of the half-filled trimer solid
(Compare the left and right sides of Figs.\ref{fig4}(c) and \ref{fig4}(d)).
As is widely accepted, half-filled electron systems is unstable
against  Mott insulator phase in the presence of electron-electron correlation.
We propose the normal state of BaIrO$_3$ is essentially 
in the Mott insulator phase, and name this ``trimer Mott insulator.''

Let us discuss here about the charge transport above $T_c$.
According to the Landau theory of metal-insulator transition
\cite{kotoliar2000}, there is a critical end point above which
metal and insulator cannot be distinguished.
In fact, the transport properties above the critical end point
can be viewed as either bad metal or bad insulator,
as reported in V$_2$O$_3$ \cite{limelette2003}
and $\kappa$-(BEDT-TTF)Cu[N(CN)$_2$]Cl \cite{kagawa2004}.
We propose that  the normal state of BaIrO$_3$ should
be understood basically in terms of trimer Mott insulator,
with the critical end point below $T_c$.
If so, the charge transport seems to show intermediate properties
between metal and insulator, which are exemplified by
the small activation energies of $E_g^{\rho}$ and $E_g^S$.
Zhao \textit{et al}. \cite{zhao2008} have found that external pressure increases
the resistivity to enhance the insulating nature of the title compound. 
They further found that the transition temperature decreases 
with pressure.
Kida \textit{et al.} \cite{kida2010} have also found
the large pressure dependence of $T_c$ in the magnetization measurement.
Their results indicate that the pressure stabilizes the insulating normal
state,  being consistent with our trimer-Mott-insulator picture above $T_c$.
Cao \textit{et al.} \cite{cao2004} have also stated 
that BaIrO$_3$ is in the verge of a metallic state.

Figure \ref{fig4}(c) schematically shows the trimer Mott insulator phase
above $T_c$.
The Ir1--Ir2--Ir1 trimer is in the higher level because of the shorter
intra-trimer Ir--Ir bond length.
All the timers have one electron in the NB $a_{1g}$ or $e''$ level,
and the energy difference $\Delta \varepsilon$ is too small to
cause charge transfer.
As the intra-trimer Ir--Ir bond length 
of the Ir3--Ir4--Ir3 trimer increases with decreasing temperature, 
$\Delta \varepsilon$ is expected to increase. 
At a certain temperature, a rapid increase of $\Delta \varepsilon$
and a charge transfer simultaneously happen,
and the electronic states  schematically drawn in Fig. \ref{fig4}(d)
realize below $T_c$,
where the NB orbital of the Ir1--Ir2--Ir1 trimer is empty 
and that of the Ir3--Ir4--Ir3 timer is fully occupied.
From the view point of valence change, the Ir$^{4+}$ state
changes into a disproportionate state of Ir$^{3.67+}$ and Ir$^{4.33+}$
in the neighboring trimers.
We regard this state as a kind of charge ordered state rather than
CDW state, because this has no relation to the Fermi surface information
such as nesting or $2k_F$ \cite{gruner1988}.
\textit{We therefore propose that the transition at 180 K in BaIrO$_3$ is 
a phase transition from the trimer Mott insulator phase to
the trimer charge ordered phase}.

We will try a rough estimate for the formal Ir valence $\nu$ 
using the bond valence sum calculation \cite{bvs} given by
$\nu = \sum_{i=1}^6 \exp[(d_0-d_i)/0.37]$,
where $d_i$ is the Ir-O bond length in a IrO$_6$ octahedron in 
units of angstrom, and $d_0 = $ 1.87 \AA\  for Ir$^{4+}$
(\url{http://www.iucr.org/resources/data/datasets/bond-valence-parameters}).
Since the positions of oxygen atoms are not well determined in the
present refinement, 
we  have instead employed the atomic position data
at 300 K from \cite{siegrist1991} and have calculated $\nu$ 
as listed in Table 1.
The 300-K data indicate that the two trimers already have different
charges;
the Ir1--Ir2--Ir1 trimer has a net charge of 12.23, whereas 
the Ir3--Ir4--Ir3 trimer has a 0.45-shorter value of 11.78.
Unlike the related ruthenate Ba$_4$Ru$_3$O$_{10}$
\cite{igarashi2013},
the charge is not disproportionated within trimer significantly. 
A slight difference (0.16 for Ir1 and Ir2; 0.07 for Ir3 and Ir4) 
may correspond to the bonding
orbital widely spread within trimer, as is pointed out in 
\cite{khomskii2012}.
For the calculation of the 16-K values, 
we used our data of $d_{12}$ and $d_{34}$ shown in 
Fig. \ref{fig3}(e), and assume that the trimers uniformly 
expand or shrink with temperature.
Namely we assume that  the Ir-O bond lengths change with temperature
in the same ratio as the Ir-Ir bond lengths.
For the Ir3--Ir4--Ir3 trimer, $d_{34}$ increases from 2.615 
to 2.640 \AA\ from 300 down to 16 K,
and thus the Ir-O bond lengths at 16 K equals
the 300-K value multiplied  by a factor of 
2.640/2.615 = 1.0096,
while they remain unchanged  for 
the Ir1--Ir2--Ir1 trimer.
With this assumption, 
$\nu$ for Ir3 and Ir4 is further reduced from the values at 300 K,
as was discussed in the last paragraph of Section 3.
At 16 K,
the Ir1--Ir2--Ir1 trimer has a net charge of 12.23, whereas 
the Ir3--Ir4--Ir3 trimer has a 1.05-shorter value of 11.18.
This is in a reasonable agreement with the picture in Fig. \ref{fig4}(d)
in spite of our rough and bold assumption.

\begin{table}[tb]
 \caption{
 Estimation of the formal Ir valence.
 The atomic positions for 300 K are employed from \cite{siegrist1991}.
 For the data at 16 K, we assume that Ir-O bond length 
 changes with  the same ratio as $d_{12}$ and $d_{34}$.
 Namely, we use the same Ir-O bond lengths as at 300 K for Ir1 and Ir2,
and values 0.96\%-larger than at 300 K for Ir3 and Ir4.
}
 \begin{center}
  \begin{tabular}{l c c c c}
   Temperature & Ir1& Ir2& Ir3& Ir4 \\
   \hline
   300 K& 4.13&  3.97& 3.95& 3.88 \\
   16 K  & 4.13&  3.97& 3.75& 3.68 \\
  \end{tabular}
 \end{center}
\end{table}

Here let us compare our model with band calculations
reported previously \cite{whangbo2001,maiti2006,ju2013}.
Whangbo and Koo \cite{whangbo2001}
have pointed out that the density of states of Ir2 and
Ir4 (the central Ir ion in the trimer) show pseudo gap at the Fermi
energy.
This is explained in terms of molecular orbital. 
The highest  occupied states belong to the non-bonding orbital of the 
$a_{1g}$ or $e''$ state of the Ir ion,
which has a node at the center of the trimer.
Note that the non-bonding orbital is expressed by 
$( \varphi_L -\varphi_R)/\sqrt{2}$, 
where $\varphi_L$ and $\varphi_R$
are the $a_{1g}$ or $e''$ state at the left and right edges of the
trimer.
The hybridization effects are also seen in other calculations
\cite{maiti2006,ju2013},
and in particular, the calculation
with strong spin-orbit interaction \cite{ju2013}
gives a small Mott gap
which is consistent with the observed small activation
energies of $E_g^{\rho}$ and $E_g^S$.
Since the band calculation was based on the structure at room
temperature, the calculated insulating states should be those above $T_c$.

Our proposed model can explain various experimental 
results at least semi-quantitatively.
It explains 
bad metal/bad insulator behavior above $T_c$, 
a charge gap below $T_c$ without translational-symmetry breaking,
and the transition susceptible to external pressure.
Since one electron in the NB $a_{1g}$ or $e''$ state per trimer
is responsible for magnetism, we can understand
the small effective magnetic moment of 0.16$\mu_B$ per Ir
above $T_c$ \cite{cao2000,kida2008}.
The transition is basically an insulator-insulator transition,
which can give anomalous critical exponents \cite{kida2008}.
This transition looks similar to the neutral-ionic transition in
the charge transfer organic complexes 
showing non-linear conduction \cite{torrance1981,tokura1988}.
BaIrO$_3$ also shows giant non linear conduction at low 
temperature \cite{nakano2006},
the origin of which may be compared with the organic complexes.
Within our model, however, the weak ferromagnetism is difficult to understand,
for all the trimers are nonmagnetic below $T_c$.
The ideal charge transfer of $e$ in Figs. \ref{fig4} (c) and (d)
may not occur in real materials, but only a fraction of $e$
may move between the trimers.
In such a case, all the trimers can be slightly magnetic and may show 
ferromagnetism with the much smaller moment of 0.016$\mu_B$
below $T_c$.

It would be fair to point out  controversial points and
limitations of our model.
First, Figure 4 is an oversimplified picture 
where transfer energies except for the intra-timer case are neglected.
In a real material, the complicated band structure may invalidate our
model, although the band calculations reported so far do not
always explain the experimental results satisfactorily.
Secondly,  the strong correlation, i.e., the on-site Coulomb repulsion
forces the electron density as homogeneous as possible, and 
disfavors the charge transfer from one trimer to another.
The proposed charge order may come from a delicate balance among
the on-site Coulomb repulsion, the energy difference between the two
timers, and the transfer energies.
A more complete theoretical modeling is needed to step further.
Third, IrO$_6$ octahedra are distorted so that there are 
no pure $a_{1g}$ and $e_g$ orbitals in a strict sense.
Accordingly the molecular orbital such as NB $a_{1g}$ may be misleading.
Even in such a case, one hole on Ir$^{4+}$ of $(5d)^5$ occupies 
the non-degenerate, highly occupied orbital (say $\phi_i$ for Ir$i$). 
Then we can make linear combinations from $\phi_1$ and $\phi_2$
for one Ir trimer, from $\phi_3$ and $\phi_4$ for the other trimer. 
Again we can construct a one electron in a corresponding
non-bonding-like state for one trimer, and 
find the essence of our model left unchanged. 

\begin{figure}[t]
 \centering
 \includegraphics[scale=.40]{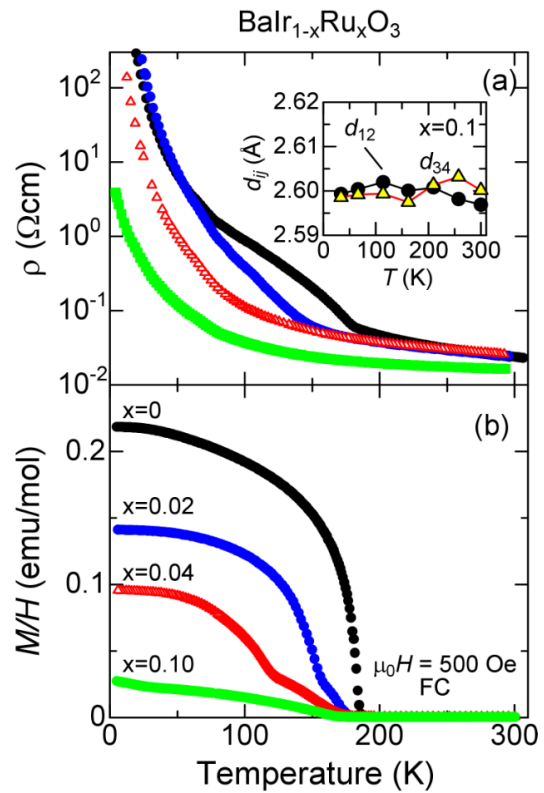}
 \caption{
The impurity effect on the 182-K transition.
(a) Resistivity $\rho$ of polycrystalline samples of BaIr$_{1-x}$Ru$_x$O$_3$.
The inset shows the intra-trimer Ir-Ir bond lengths for the $x=0.1$ sample.
The two trimers have the identical, temperature-independent bond length.
(b) Magnetization $M$ divided by an external field $H$
of 500 Oe of the same samples.
}
 \label{fig5}
\end{figure}

To examine the validity of our model, we have investigated 
the impurity effects on the 182-K transition for a set of
polycrystalline samples of BaIr$_{1-x}$Ru$_{x}$O$_3$.
Figure \ref{fig5} shows the magnetization and the resistivity of the
samples. 
The phase transition is systematically and drastically suppressed with
the Ru concentration $x$, and essentially vanishes at
only 10\% substitution;
For $x=0.1$, the resistivity is non-metallic without any trace 
of anomalies at all temperatures measured,
whereas a tiny trace of the weak ferromagnetic phase of $x=0$
is visible in the magnetization measurement.
The resistivity for $x=0.1$ indicates the electronic states above $T_c$
is essentially insulating, being consistent with the picture of 
the trimer Mott insulator.
The inset of Fig. \ref{fig5}(a) shows the intra-trimer 
Ir-Ir bond lengths for the $x=0.1$ sample, 
where the two lengths of $d_{12}$ and $d_{34}$ are
essentially temperature-independent, and identical.
Concomitantly the negative expansion of the $c$-axis
is severely suppressed (not shown).
The Ru substitution simultaneously suppresses the 182 K-transition  
and the electron transfer from Ir1--Ir2--Ir1 to Ir3--Ir4--Ir3,
and therefore these two phenomena are two sides of the same coin.
We further note that the charge order is known to be 
vulnerable to impurities \cite{tatami2007}, 
while the charge density wave is more or less 
robust against impurities \cite{kawabata1985}.

A similar type of the phase transition occurs in 
the organic salt 
$\beta$-(meso-DMBEDT-TTF)$_2$PF$_6$ \cite{kimura2006}.
Many conductive organic salt with a chemical formula of
$A_2B$ consists of 
a dimer molecule of $A_2$ as a composition unit.
In the present case,
two molecules of meso-DMBEDT-TTF form a dimer structure,
and share one hole per dimer to behave as
\textit{dimer Mott insulator} \cite{kino1995}.
Okazaki \textit{et al}. \cite{okazaki2013}
identified the phase transition in this compound
at 70 K to a transition from dimer Mott insulator phase to
charge ordered insulator phase.
This results from a competition 
between the intra- and inter-dimer Coulomb interactions,
which causes intrinsic inhomogeneous state detected through
infrared microscope measurement.
As is similar to BaIrO$_3$, this organic salt also shows 
non-linear conduction below $T_c$  \cite{niizeki2008}.

Finally we briefly comment that we have applied
a similar molecular orbital concept to the related timer Ru oxide 
Ba$_4$Ru$_3$O$_{10}$ in
which the three face-shared RuO$_6$ octahedra form a Ru$_3$O$_{12}$
trimer \cite{klein2011,igarashi2013,igarashi2015}.
This related oxide shows a phase transition at 105 K,
where a metallic, paramagnetic state changes into
an insulating, antiferromagnetic state at low temperature.
However, the electronic states are different;   
In this related oxide, each trimer is  three electrons short because
Ru$^{4+}$ has four $4d$ electrons.
On the basis of a similar molecular orbital approach,
we have to consider partially occupied $e_{g}'$ orbitals.
In this case, the static electronic potential is deeper at the center
of the trimer, and a finite fraction of charge is expected
to transfer from the edge.
Namely, intra-timer charge transfer drives the phase transition
in Ba$_4$Ru$_3$O$_{10}$.
This makes a stark contrast to the inter-trimer charge transfer 
in BaIrO$_3$.
The intra-trimer charge disproportionation has been
also discussed from the local-density-approximation calculations
\cite{khomskii2012,radke2013}.

\section{Summary}
In summary, we have prepared polycrystalline samples of the trimer oxide
BaIrO$_3$ with the face-shared Ir$_3$O$_{12}$ trimers,
and have measured the resistivity, thermopower, magnetization and
synchrotron x-ray diffraction from room temperature down to 4 K.
On the basis of the structure analysis, we have made a model for
the electronic states of this oxide, and have pointed out that
this oxide can be regarded as a kind of Mott insulator,
when the Ir$_3$O$_{12}$ trimer is identified to one lattice site.
Within the same framework, the 182-K transition can be 
viewed as a transition from the trimer Mott insulator phase to
the trimer charge-ordered phase,
which is essentially different from conventional charge density wave
as was used to be a candidate for the transition.
We have further succeeded in explaining 
various properties above $T_c$ such as 
the non-metallic resistivity
and  thermopower, the large pressure effect,
the unusually small effective magnetic moment.
We expect that the proposed concept of trimer solid
will  be applied to other 
face-shared transition metal oxides.

\begin{acknowledgments}
 The authors would like to thank Mike Whangbo for showing 
unpublished data for the electronic states of BaIrO$_3$.
They are indebted to T. Nakano, T. Kida, M. Hagiwara, Y. Nogami, S. Mori 
for collaboration at an early stage of this work.
They also wish to appreciate J. Akimitsu, T. Arima, S-W. Cheong,
M. Tsuchiizu for fruitful discussion and useful advise.
This work was partially supported by 
Grant-in-Aid for Scientific Research, Japan Society for the  
Promotion of Science, Japan 
(Kakenhi Nos. 25610091, 26247060),
and by Program for Leading Graduate Schools ``Integrative Graduate
Education and Research in Green Natural Sciences'', MEXT, Japan.
The synchrotron x-ray diffraction was performed under
the approval of the Photon Factory Program
Advisory Committee (Proposal Nos. 2012G718 and 2012S2-005).
\end{acknowledgments}

\bibliography{ito}
\bibliographystyle{apsrev4-1}
\nocite{*}

\end{document}